\documentclass{eptcs}
\providecommand{\event}{TFPIE 2021}

\usepackage{mweights}
\usepackage{fontaxes}
\usepackage{graphicx}
\usepackage{caption}
\usepackage{subcaption}
\usepackage{stfloats}
\usepackage{multido}
\usepackage{setspace}
\usepackage{textcomp}
\usepackage{amssymb}
\usepackage{amsmath}
\usepackage{alltt}
\usepackage{tikz}
\usepackage{pgfplots}
\RequirePackage{doi}
\usepackage{cleveref}
\usepackage{hyperref}

\newcommand{\elist}{\texttt{\textquotesingle{()}}}
\newcommand{\arrow}{\(\rightarrow\)}
\newcommand{\dotss}{\(\ldots\)}
\newcommand{\vdotss}{\(\vdots\)}

\newcommand{\htdp}{\texttt{HtDP}}

\newcommand{\racket}{\texttt{Racket}}
\newcommand{\oo}{\texttt{OO}}
\newcommand{\oop}{\texttt{OOP}}

\newcommand{\dsl}{\texttt{DSL}}
\newcommand{\comment}[1]{}

\title{Design of Classes I}

\author{Marco T. Moraz\'an
\institute{Seton Hall University}
\email{morazanm@shu.edu}}
\def\titlerunning{Design of Classes I}
\def\authorrunning{M. T. Moraz\'an}

\begin{document}
\maketitle

\begin{abstract}
The use of functional programming languages in the first programming course at many universities is well-established and effective. Invariably, however, students must progress to study object-oriented programming. This article presents how the first steps of this transition have been successfully implemented at Seton Hall University. The developed methodology builds on the students' experience with type-based design acquired in their previous introduction to programming courses. The transition is made smooth by explicitly showing students that the design lessons they have internalized are relevant in object-oriented programming. This allows for new abstractions offered by object-oriented programming languages to be more easily taught and used by students. Empirical evidence collected from students in the course suggests that the approach developed is effective and that the transition is smooth.
\end{abstract}

\section{Introduction}
\label{intro}
Designed based introductory courses to programming, inspired by the approach put forth in the textbook \emph{How to Design Programs} (\htdp) \cite{HtDP,HtDP2}, are now well-established and effective \cite{rainfall,mtm27,bootst}. This approach focuses on problem solving using type-based program design. That is, solutions to problems are designed based on the data that needs to be processed. One of its most salient features is the use of \emph{design recipes}. A design recipe is a series of steps that take a programmer from a problem statement to a working and tested solution. Each step has a concrete outcome that students and instructors can verify. This approach forces students to focus on problem analysis, data representation, and testing. Mistakenly, many believe that the approach is limited to problems solved using structural recursion. Although a design recipe facilitates designing programs using structural recursion, there are also design recipes, albeit less prescriptive, for programs based on generative and accumulative recursion \cite{HtDP,HtDP2}, distributed programming \cite{mtm27}, and imperative while loops \cite{mtm30}. That is, a design recipe is more about problem solving than it is about prescriptive steps for a given type instance to be processed. Nonetheless, the pivotal role of types give rise to different (not all) design recipes.

Students that start with a design-based curriculum using a functional programming language eventually must move on to other courses that introduce them to an object-oriented (\oo) programming language. How is this effectively done? How do we move beginners from program design using a functional programming language to program design using an \oo \ programming language? They keyword here is \emph{design}. We certainly should not have students migrate to writing programs without designing them. This article presents how this transition starts at Seton Hall University within the context of the first \oo-based programming course. At the heart of the approach is type-oriented programming and the use of design recipes to guide students during development. The goal is introduce students to abstractions offered by \oo \ programming languages without starting from scratch and to make a smooth transition from using a functional programming language to an \oo \ programming language. That is, our aim is to build on and reinforce the program design skills that students have internalized. A natural question that may arise in the mind of some readers is: Why run another course using design recipes if students are already familiar with design recipes? The answer is twofold. First, design skills do not magically transfer from functional programming to object-oriented programming (\oop) automatically in beginners. It is important to show beginning students how to apply their design skills in a new context. Otherwise, it is all too easy for beginning students to set aside the lessons of design and resort to trial-and-error hacking that leads to programs that students cannot convincingly argue are correct and that instructors have difficulty comprehending. Second, the use of design recipes brings \oop \ to familiar territory for students. Thus, facilitating a smooth transition to \oop \ and giving students a strong sense that lessons learned in previous courses are relevant and useful when using a programming language that is popular in industry.

This article outlines how students are introduced, at the beginning of their first \oop \ course, to objects, interfaces, generic programming, polymorphic dispatch, inheritance, $\lambda$-expressions, and abstract classes using \texttt{Java}\footnote{Latter parts of the course cover mutation, arrays, \texttt{ArrayList}s, \texttt{while}-loops, design patterns, iterators, and hash tables.}. Unlike most textbooks, the emphasis is not on syntax. Instead, the emphasis is to build on the design experience students have accumulated and bring to the course. The article is organized as follows. \Cref{rw} discusses related work. \Cref{sb} presents the students' background. \Cref{oopintro} presents how classes are motivated using the students' experience with structures. \Cref{utypes} motivates interfaces, generic programming, polymorphic dispatch, inheritance, and $\lambda$-expressions building on students' design experience with union types. \Cref{abstraction} motivates abstract classes based on students' experience with functional abstraction. \Cref{feedback} presents student feedback. Finally, \Cref{fw} draws conclusions and presents directions for future work.

\section{Related Work}
\label{rw}

Broadly speaking, there are two approaches to introduce beginners to \oop. The most closely related to the approach described in this article are those approaches that teach design principles using a \dsl. The other approach is more widely spread and is based on teaching the syntax of a language. There are strong advocates in both camps each with valid points to make. The advocates for teaching design put forth that design principles are language agnostic and may be applied regardless of the syntax used. Currently, a major problem with this approach is the lack of textbooks available that students and instructors may follow. The advocates for a syntax-based approach put forth that there is a long history of teaching syntax and numerous textbooks that students and instructors may rely on. Furthermore, they observe that most new faculty members face a steep learning curve when adopting a design-based approach. This article will not settle this debate, but it is fairly clear that a popular syntax today is left behind when a new programming language becomes fashionable. Therefore, an emphasis on design rather than syntax is likely to be more beneficial to students especially in the long term.

\subsection{DSL-Based Approaches}

The unpublished textbook \texttt{How to Design Classes} (\texttt{HtDC}) is closely coupled with a \dsl \ known as \texttt{ProfessorJ} \cite{ProfJ}. It first introduces students to classes as a mechanism to define a type of data that may have many fields just like a structure may have many fields. A field type in a class, unlike a field type in a \htdp \ structure, must be part of the program's code. As a structure, a class has a constructor that initializes the fields and produces an object--an instance of the type defined by the class. This class constructor is explicitly written by the programmer. The design of a class starts by understanding how data is to be represented. Union types are introduced to motivate the need for interfaces. In essence, an interface is used to define a type and glue together its subtypes. Interfaces are particularly useful to define data of arbitrary size (e.g., a list) and, therefore, programs based on structural recursion. The design of methods is not tackled until students develop some expertise defining classes that only have a single constructor. The design of a method follows the design recipe approach found in \htdp. Testing of methods is done using syntax as the following:
\begin{alltt}
     check item.price() expect 31
\end{alltt}
\texttt{HtDC} has students engage in problem and data representation analysis before developing methods. Furthermore, students are encouraged to write tests before writing a method. Similar to \texttt{HtDC}, the approach described in this article starts with parallels between structures and classes with no subtypes and then moves to union types and abstraction. In contrast, however, we start directly with \texttt{Java} syntax and rely quite heavily on the knowledge students acquired designing functions using a functional programming language. For example, students have been introduced to generic types and these are introduced early in the course. In this manner, for instance, students reason and design based on a list of \texttt{X} instead of a list of geometric shapes, a list of points, or a list of strings.

Another design-based approach developed at Northeastern University teaches students (that have used \htdp \ in their first course) using a hybrid approach \cite{PwC}. The beginning of the second course proposes the use of several \dsl~s \ embedded in \racket \ before making the full transition into \texttt{Java}. The motivation for such an approach is that students first develop a command of basic \oop \ principles before tackling \texttt{Java}'s syntax, types, and a compiler (as opposed to an interpreter). This approach also relies heavily of the design recipe to provide students (and instructors) with a framework for program design and discussion. From the start objects are described as consisting of data and functions. A class is first thought of as a structure and its fields are accessed using a dot notation. For example, \texttt{(point . x)} denotes that the message, \texttt{x}, on the right hand of the dot is used to request a value from the object, \texttt{point}, on the left hand side of the dot. Union types and design based on structural recursion are introduced as requiring a distinct class for each variety in the union. The transition to \texttt{Java} occurs in the middle of the semester. This transition is made to functional \texttt{Java} first to make programs as similar to those developed using the \texttt{DSL}s embedded in \racket. Later mutation and looping constructs are presented. In a similar spirit, the work presented in this article introduces objects as data and functions. Objects are introduced using functional \texttt{Java} to have students focus on the principles of \oo \ design instead of properly making mutations. In contrast, the work described here is presented to students after two semesters (not one) designing using a functional programming language. It relies much more on student programming-maturity that allows them to tackle a new syntax (i.e., \texttt{Java}) from the beginning of the course. Furthermore, this experience allows us to quickly and smoothly introduce students to generic programming, interfaces, $\lambda$-expressions, and abstract classes.

\subsection{Java-Based Approaches}

Most \oop \ textbooks for beginners (and, therefore, probably most instructors) follow a syntax-based approach (e.g., \cite{Ford,Goodrich,Sedgewick,Tymann}). The expectation is that somehow students will learn how to successfully write \oo \ programs based on syntax, examples, and in-class warnings about common pitfalls. Although many readers of this article, surely, ``learned" how to program in this manner, it is not a stretch to say that \oo \ abstractions are not effectively taught in this manner. Simply stated, programmers need to understand what an abstraction represents and how it was developed. This kind of insight is unlikely to be discovered by novices learning syntax by example. In addition, beginning students need a systematic way to design large software systems--one of the principle goals of \oop. Students must learn to design a (hierarchical) class system and understand how components in a system are expected to interact. In contrast to this approach, the work described in this article motivates the existence of abstractions (such as classes, interfaces, and abstract classes) and how to design programs using them.

\texttt{DrJava} is designed to gently introduce students to writing \texttt{Java} programs and to provide support for developing programs \cite{DrJava}. It is an \texttt{IDE} that provides students with a definitions and an interactions pane that hide the details of interacting with a \texttt{Java} compiler. Students are exposed to full \texttt{Java} syntax from the beginning including types, classes, and methods. In addition, \texttt{DrJava} provides support for interactions with error messages. Like \texttt{DrJava}, the approach described here has students use \texttt{Java} syntax from the start. Unlike the \texttt{DrJava} approach, we allow students to directly start wrestling with the error messages of an industrial compiler. This is based on the observation that the work described here takes place in their third programming course. Therefore, there is more programming maturity among our students than among beginners in a first semester course.

\texttt{BlueJ} is an \texttt{IDE} specifically developed to teach \oop \ to beginners \cite{BlueJ}. Its interface shields programmers from interacting directly with the compiler and presents beginners with a main window displaying a class diagram to visualize the application's structure. Programmers can then, for example, click on a class to edit it. Its pedagogy starts with objects first, but never with a blank screen. Students read relatively large \texttt{Java} code from the beginning and are asked to modify it. As already mentioned above, like \texttt{BlueJ}, the work presented here has students work directly with \texttt{Java} from the beginning. In contrast, students are trained to design classes from the beginning instead of working with an unfamiliar design. This decision was made because classes are central in \oo \ design. Methods inside a class only make sense if the overall design of the system is understood. That is, methods only make sense if it is understood why they are encapsulated in a given class.

At Northeastern University the course \emph{Fundamentals II Introduction to Class-based Program Design} uses \texttt{Java} in conjunction with \htdp-inspired design recipes \cite{Lerner}. As the approach presented in this article students are taught systematic design from the beginning using \texttt{Java}. Among other things, this means that contracts immediately become part of the code to define the types of variables and methods. Another similarity is that students are quickly introduced to testing, interfaces and abstract classes. In contrast, tests for methods to access fields are not introduced. The work presented emphasizes these because beginning students eagerly insist in not writing such methods which becomes a problem when fields become private. Another contrasting feature is that the Northeastern course does not introduce students to $\lambda$-expressions in \texttt{Java}. As $\lambda$-expressions become part of idiomatic \texttt{Java} it is important to introduce students to their use and the convenience they provide.

\section{Student Background}
\label{sb}
At \texttt{SHU}, the first two introductory Computer Science (\texttt{CS}) courses focus on problem solving using a computer \cite{mtm22,mtm28}. The languages of instruction are the successively richer subsets of \texttt{Racket} known as the student languages which are tightly-coupled with \texttt{HtDP} \cite{HtDP,HtDP2}. No prior experience with programming is assumed. The first course starts by familiarizing students with primitive data (e.g., numbers, strings, booleans, symbols, and images), primitive functions, and library functions to manipulate images. During this introduction, students are taught about variables, defining their own functions, and the importance of writing signatures, purpose statements, and unit tests. Through the development of signatures students are introduced to types albeit as comments in their code. The course then introduces students to data analysis and programming with compound data of finite size (i.e., structures). At this point, students are exposed to their first design recipe. Students gain experience in developing data definitions, type instances, function templates, and tests for all the functions they write. Building on this experience, students develop expertise with processing compound data of arbitrary size such as lists, natural numbers, and trees. In this part of the course, students learn to design functions using structural recursion. After structural recursion, students are introduced to functional abstraction and the use of higher-order functions such as \texttt{map} and \texttt{filter}. This part of the course also introduces students to generic types as these are necessary to write contracts for higher-order functions. The first semester ends with a module on distributed programming \cite{mtm26,mtm27}.

In the second course, students are exposed to generative recursion, accumulative recursion, and mutation \cite{mtm28}. The course starts with generative recursion and transitions to accumulative recursion. The course then introduces students to mutation. This part of the course includes a module on interfaces as a mechanism to encapsulate state variables and functions on these state variables. These functions are called constructors (to build a new instance of the type), observers (to compute a value from a given type instance), and mutators (to assign a new value to a state variable). Message-passing is used to request a service from an interface. In essence, an instance of an interface is an object. This part of the course also includes a module on \texttt{while} loops \cite{mtm30}. This module introduces students to the proper sequencing of mutations to obtain a computational goal using loop invariants and Hoare Logic \cite{Hoare,Hoare2}.

Each course is for a semester (15 weeks). There are two weekly 75-minute lectures that students are required to attend. The typical classroom has between 15 to 25 students. In addition to the lectures, the instructor is available to students during office hours and there are about 30 hours of tutoring each week which the students may voluntarily attend. The tutoring hours are conducted by undergraduate students handpicked and trained by the lecturer.

The use of different type-based design recipes is emphasized throughout both courses. We can outline these design recipes as follows:
\begin{alltt}
     1. Problem Analysis                    5. Write Function Header
     2. Data Analysis                       6. Write Unit Tests
     3. Function Template Development       7. Write Function Body
     4. Signature and Purpose Statement     8. Run Tests
\end{alltt}
Students first outline how to solve a problem. They then proceed to create data definitions for the required data's representation (i.e., they define types). Based on the type, a function template is developed that outlines how to process instances. After these initial steps, problem-specific steps follow. Students start with a signature, a purpose statement, and a function header. Armed with these students proceed to write unit tests before writing the function's body. Finally, students run the unit tests and redesign if necessary.

\section{From Structures to Classes}
\label{oopintro}

Students arrive to the course having familiarity with structures and the design of functions to process a structure. This type of data does not have varieties (i.e., no subtypes). In other words, it is not a union type. We leverage this knowledge to motivate the need for classes.

\subsection{Students' Structure Knowledge}
\label{stud}
\begin{figure}
\centering
\begin{alltt}
     ;; student is a structure
     ;;     (make-student string 0\(\leq\mathbb{R}\leq\)4.0 natnum)
     ;; with a name, a grade point average, and a number of completed credits.
     (define-struct student (name gpa credits))
     ;; Sample students
     (define STUDENT1 (make-student "Spiderman" 3.5  16))	
     (define STUDENT2 (make-student "Batwoman"  3.9  43))
     (define STUDENT3 (make-student "Superman"  3.7  75))	
     (define STUDENT4 (make-student "Ironman"   4.0  120))
     ;; student \(\rightarrow\) string		
     ;; Purpose: Determine the given student's year level
     (define (student-level a-student)
       (cond [(<= (student-credits a-student) 30) "Freshman”]
             [(<= (student-credits a-student) 60) "Sophomore”]
             [(<= (student-credits a-student) 90) "Junior”]
             [else "Senior”]))
     (check-expect (student-level STUDENT1) "Freshman”)
     (check-expect (student-level STUDENT2) "Sophomore”)
     (check-expect (student-level STUDENT3) "Junior”)
     (check-expect (student-level STUDENT4) "Senior”)
\end{alltt}
\caption{Function to Determine a Student's Year Level} \label{student}
\end{figure}

To illustrate a student's background consider solving the following problem:
\begin{alltt}
     A student has a name, grade point average, and the number of credits
     completed. Write a function that takes as input a student and that
     returns the given \texttt{student}'s year level: Freshman, Sophomore,
     Junior, or Senior.
\end{alltt}
As part of the first step of the design recipe students outline how to solve the problem as follows:
\begin{alltt}
     A student with 30 credits or less is a freshman
     A student with 31-60 credits is a sophomore
     A student with 61-90 credits is a junior
     Otherwise, the student is a senior
\end{alltt}
This analysis already informs the student that a conditional expression is required to solve this problem.

Given that a \texttt{student} is compound data of finite size it may be represented using structure:
\begin{alltt}
     (define-struct student (name gpa credits))
\end{alltt}
The structure has 3 fields: a string for the name, a nonnegative real number for the grade point average, and a natural number for the number of completed credits. Once a data definition and structure definition are written students define sample instances. In this example, at least four are required given that a \texttt{student} may be at one of four levels. Students know that the above structure definition creates a constructor (\texttt{make-student}) and four observers (\texttt{student-name}, \texttt{student-gpa}, \texttt{student-credits}, and \texttt{student?})\footnote{They are also aware that a mutator for each field is created.}.

The function template for Step 3 must capture all the commonalities that all functions that process a \texttt{student} may have. This step is problem-agnostic and may be used (as the result of Step 2) to solve other problems that require processing a \texttt{student}. A \texttt{student}-processing function needs at least a \texttt{student} as input. Given that a student structure has three fields there are 3 selector expressions in the body of the definition template that may be useful. In addition, there must be at least one test given that there is only one variety of \texttt{student} (i.e., there are no subtypes). The template for a function on a \texttt{student} is:
\begin{alltt}
     ;; student \dotss  \arrow \dotss     Purpose: \dotss
     (define (f-on-student a-student)
       \dotss(student-name a-student)\dotss(student-gpa a-student)\dotss
       \dotss(student-credits a-student)\dotss)
     (check-expect (f-on-student \dotss) \dotss) \dotss
\end{alltt}

The solution with the rest of the design recipe steps completed is displayed in \Cref{student}. The output of the function may be represented as a string (e.g., \texttt{"Freshman"}). We briefly observe that, as expected, the function returns a string, the body of the function is a conditional expression, and there is one test for each possible outcome\footnote{Technically, the function returns an instance of an enumerated type (containing 4 strings), but such details are omitted here in the interest of brevity.}.

\subsection{Classes and Objects}
\label{objects}

Students are explained that a type may be described as data and the operations that are valid on instances of the type. Code is organized in a class and instances of the class are called objects. Unlike structures, however, the functions (i.e., methods) implementing the valid operations must all be written by the programmer. In addition, unit tests are written in a separate file.

This means that the basic design recipe must be updated to encompass this perspective. Students are explained that the methods implementing the valid operations on the type being defined must all be encapsulated in the class along with the data that may vary from one instance to the next. The following design recipe is presented:
\begin{enumerate}
  \item Problem Analysis
  \item Data Analysis
  \item Class Template Development
  \item Tests Template Development
  \item Write Tests
  \item Method Development
    \begin{enumerate}
      \item Write Signature and Purpose Statement
      \item Write Method Header
      \item Write Unit Tests
      \item Write Method Body
    \end{enumerate}
  \item Run Tests
\end{enumerate}
Problem analysis remains unchanged. Data analysis now uses an object, instead of a structure, to describe compound data of finite size. The next two steps outline the basic structure of any class and testing class for the type being defined. Once students have a testing template they develop tests for the valid operations. Method development for the valid operations proceeds in a manner similar to functions using a functional programming language. A major difference is that signatures (i.e., types) are part of the code and not comments. Finally, students must run their tests and redesign if necessary. Observe that each step of the design recipe has a specific outcome that may be verified by a student or instructor reading the code.

To make the steps of the design recipe concrete let us revisit the student problem from \Cref{stud}. Problem analysis remains unchanged. That is, a \texttt{student}'s year level is determined the same way. Data analysis defines a \texttt{Student} as an object:
\begin{alltt}
     Student is an object, new student(String 0\(\leq\mathbb{R}\leq\)4.0 natnum),
     with a name, a grade point average, and a number of completed credits.
\end{alltt}
The following operations are valid on a student:
    \begin{description}
      \item[Constructors]
        \begin{description}
          \item[]
          \item[Student] String double int \arrow \ Student
        \end{description}
      \item[Observers]
        \begin{description}
          \item[]
          \item[getName]:\ \ \ \arrow \ String
          \item[getGpa]: \ \ \ \ \ \arrow \ double
          \item[getCredits]: \arrow \ int
          \item[isStudent]:\ \ \ \arrow \ boolean
        \end{description}
    \end{description}
Observe that the description of the operations includes the signatures for each method. For example, \texttt{isStudent} takes no input and returns a \texttt{boolean}. It is noteworthy that this constant \texttt{true} method is included to have students develop the habit of developing predicates which become useful when there is variety in the data. In addition, it is consistent with student expectations moving from \racket \ structures to classes.

Class template development uses the signatures from the previous step to outline any class that implements \texttt{Student}. The template must include instance variables and method signatures. At this point we do not yet separate specification from implementation. This allows students to more easily develop classes. For our \texttt{Student} example, therefore, the class template is\footnote{In the interest of saving space \texttt{Java} code formatting and naming conventions are not always followed in this article.}:
\begin{alltt}
     class Student
     \{String name;     double gpa;   int credits;
      //               0\(\leq\)gpa\(\leq\)4.0    credits\(\geq\)0
      Student(String nm, double g, int c)
      \{name = nm;     gpa = g;     credits = c; \}
       String  getName()    \{ return(name);  \}
       double  getGpa()     \{ return(gpa); \}
       int     getCredits() \{ return(credits);  \}
       boolean isStudent()  \{ return(true); \}\}
\end{alltt}
Observe that when specification is not separated from implementation, the class template contains full implementations of the valid operations. For students, this approach brings their familiarity with structures and function design to bear on class design. It is also noteworthy that \texttt{name $\neq$ null} is not listed as an invariant. This is because students have not been exposed to \texttt{null} yet.

Test template development has students outline the type instances and the tests that must be implemented. For compound data of finite size there must be at least one instance and there must be at least one test for each valid operation. For our \texttt{Student} example the testing class template is:
\begin{alltt}
    class TestStudent
    \{ @Test
      public void testStudentMethods()
     \{	Student S\(\sb{1}\) = new Student(\dotss);
             \vdotss
       Student S\(\sb{n}\) = new Student(\dotss);
       assertEquals(\dotss.getName(), \dotss);
       assertEquals(\dotss.getGpa(), \dotss, \dotss);
       assertEquals(\dotss.getCredits(), \dotss);
       assertEquals(\dotss.isStudent(), \dotss);
             \vdotss \}\}
\end{alltt}
The template states that first instances must be developed. This is followed by tests for the observers using the defined instances. At this point, students are told that \texttt{public} is part of the required syntax.

\begin{figure}
\centering
\begin{alltt}
     public class TestStudent
    \{ @Test
      public void testStudentMethods()
      \{Student S1 = new Student("Spiderman" 3.5  16);
       Student S2 = new Student("Batwoman" 3.9  43);
       Student S3 = new Student("Superman"   3.7  75);
       Student S4 = new Student("Ironman"  4.0  120);
       assertEquals(S1.getName(),    "Spiderman");
       assertEquals(S1.getGpa(),     3.5, 0.01);
       assertEquals(S1.getCredits(), 16);
       assertEquals(S1.isStudent(),  true);
       assertEquals(S1.getLevel(),   "Freshman");
       assertEquals(S2.getLevel(),   "Sophomore");
       assertEquals(S3.getLevel(),   "Junior");
       assertEquals(S4.getLevel(),   "Senior"); \}\}
\end{alltt}
\caption{Tests for Student Class} \label{studtests}
\end{figure}

The next step of the design recipe has students specialize the testing template. There are two important observations to make. The first is that at this point students are not wondering what needs to be tested. The second is that tests are written before specializing the class template. Sequencing design-recipe steps in this manner becomes useful when an actual problem is solved. Writing tests commonly provides insights into how a value may be computed. The \texttt{Student} tests may be specialized as displayed in \Cref{studtests}.

At this point students have a \texttt{Student} implementation and may run the tests to validate the implementation allowing them to now solve problems. This requires designing methods and tests that are added to the existing \texttt{Student} and \texttt{TestStudent} classes. For our year level example this requires performing Steps 5-7 of the design recipe given that problem and data analysis as well as template development have already been done. In other words, students build on the work they have done. The following tests are added for Step 5:
\begin{alltt}
     assertEquals(S1.getLevel(), "Freshman");
     assertEquals(S2.getLevel(), "Sophomore");
     assertEquals(S3.getLevel(), "Junior");
     assertEquals(S4.getLevel(), "Senior");
\end{alltt}
Students observe that these are the same tests written for the structure-based version albeit using different syntax. The following method is added to the \texttt{Student} class:
\begin{alltt}
     Purpose: Determine the given student's level
     String getLevel()
     \{if (credits <= 30) \{return("Freshman");\}
      else if (credits <= 60) \{return("Sophomore");\}
           else if (credits <= 90) \{return("Junior");\}
                else \{return("Senior");\} \}
\end{alltt}
Once again, students appreciate that in essence this is the same function as in the structure-based version albeit using different syntax. It is worth noting that students do comment on the more cumbersome conditional statement syntax.

\section{From Union Types to Interfaces}
\label{utypes}

Problem solving using union types is used to introduce students to interfaces, generic programming, $\lambda$-expressions, and polymorphic dispatch. Once again, these topics are not covered from scratch. Instead, we build on the knowledge students bring to the classroom. Specifically, we build on knowledge of parameterized data definitions and functional interfaces using message passing.

\subsection{Students' Background}

Students arrive knowing how to abstract over data definitions to obtain parameterized (or generic) data definitions in which type variables are used instead of concrete types. For example, consider the following data definitions:
\begin{alltt}
     ;; A list of number (lon) is one of: empty or (cons number lon)
     ;; A list of string (los) is one of: empty or (cons string los)
\end{alltt}
Students abstract over them to obtain the following parameterized data definition:
\begin{alltt}
     ;; A (listof X), loX, is one of: empty or (cons X loX)
\end{alltt}
Here \texttt{X} is a type variable. Given that this is a union type (i.e., the data has subtypes) students know that a conditional is needed to distinguish between the subtypes in the body of any function that processes a \texttt{(listof X)}. Such data definitions are used to write generic functions like:
\begin{alltt}
     ;; [X Y]: (X \arrow Y) (listof X) \arrow (listof Y)
     ;; Purpose: To apply the given function to the given list
     (define (mapf f a-lox)
       (if (empty? a-lox)
           \elist
           (cons (f (first a-lox)) (mapf f (rest a-lox)))))
     (check-expect (mapf add1 \elist)        \elist)
     (check-expect (mapf add1 \textquotesingle{(1 2 3)})   \textquotesingle{(2 3 4)})
\end{alltt}
\texttt{X} and \texttt{Y} are type variables whose values are unknown until \texttt{mapf} receives its arguments. Signatures are parameterized with the unknown types. Students understand that functions parameterized in this manner are generic.

\begin{figure}[t]
\begin{minipage}[b]{.5\linewidth}
\begin{alltt}
(define (mtList)
 (local
  [(define (mycons an-x)
    (consList an-x service-manager))

   (define (equals L)
     (L \textquotesingle{empty?}))


   ;; [Y]: (X \arrow Y) \arrow I(listof Y)
   (define (map f) service-manager)



   ;; [A]: (X \arrow A) A \arrow A
   (define (foldl f acc) acc)

   (define (service-manager m)
    (cond [(eq? m \textquotesingle{first})  (error \dotss)]
          [(eq? m \textquotesingle{rest})   (error \dotss)]
          [(eq? m \textquotesingle{empty?})  \#true]
          [(eq? m \textquotesingle{cons})    mycons]
          [(eq? m \textquotesingle{equals})  equals]
          [(eq? m \textquotesingle{map})     map]
          [(eq? m \textquotesingle{foldl})   foldl]
          [else (error \dotss)]))
    service-manager))
  \end{alltt}
  \subcaption{Empty-List Interface}\label{ielist}
\end{minipage}
\begin{minipage}[b]{.5\linewidth}
\begin{alltt}
(define (consList first rest)
 (local
  [(define (mycons an-x)
     (consList an-x service-manager))

   (define (equals L)
     (and (equal? first (L \textquotesingle{first}))
          ((rest \textquotesingle{equals}) (L \textquotesingle{rest}))))

   ;; [Y]: (X \arrow Y) \arrow I(listof Y)
   (define (map f)
     (consList (f first)
              ((rest \textquotesingle{map}) f)))

   ;; [A]: (X \arrow A) A \arrow A
   (define (foldl f acc)
     ((rest \textquotesingle{foldl}) f (f first acc)))
   (define (service-manager m)
     (cond [(eq? m \textquotesingle{first})   first]
           [(eq? m \textquotesingle{rest})    rest]
           [(eq? m \textquotesingle{empty?})  \#false]
           [(eq? m \textquotesingle{cons})    mycons]
           [(eq? m \textquotesingle{equals})  equals]
           [(eq? m \textquotesingle{map})     map]
           [(eq? m \textquotesingle{foldl})   foldl]
           [else (error \dotss)]))]
    service-manager))
 \end{alltt}
 \subcaption{Cons-List Interface}\label{iclist}
\end{minipage}
\caption{\texttt{Ilox} Implementation.}
\label{ilox}
\end{figure}

Students also arrive with experience designing functional interfaces using message-passing. An interface defines the operations that are valid on the type defined. As an example, consider the following small interface, \texttt{Ilox}, for \texttt{(listof X)}:
\begin{alltt}
     An IloX is an interface that provides the following services:
           first:  X throws error          rest: Ilox throws error					
          empty?: boolean                  cons: (X \arrow Ilox)									
          equals: Ilox \arrow boolean       [Y] map: (X \arrow Y) \arrow IloY
       [A] foldl: (X A \arrow A) A \arrow A
\end{alltt}
In an interface a service that requires more input returns a function that consumes that input to compute the answer. Otherwise, it returns a value. An interface definition specifies the type of the returned value. Functions like \texttt{map} and \texttt{foldl} are only parameterized with one type variable, \texttt{Y} and \texttt{A} respectively, because they are encapsulated inside the interface \texttt{ILoX}. That is, \texttt{X} is known in any instance of \texttt{ILoX}. Finally, students understand that since \texttt{(listof X)} is a union type \texttt{Ilox} must be implemented twice (once for each subtype). Sample implementations are displayed in \Cref{ilox}. Observe that students have been exposed to polymorphic dispatch. There is no need for a conditional expression, for example, to implement functions such as \texttt{foldl} or \texttt{map} because each subtype interface encapsulates its own version of these functions. Finally, we note that \texttt{equals} is for extensional equality.

\subsection{Union Types and Classes}
\label{uniontypes}

Students are presented with the following design recipe to implement union types in an \oo \ language:
\begin{enumerate}
  \item Problem Analysis
  \item Data Analysis Reveals Need for a Union Type
  \item Design Interface
  \item Develop Unit Tests for Interface
  \item Implement the Interface for each Subtype Using a Class
  \item Method Development for each Class
   \begin{enumerate}
      \item Write Purpose Statement
      \item Write Method Header
      \item Write Unit Tests
      \item Write Method Body
   \end{enumerate}
  \item Run Tests
\end{enumerate}
Students perform problem and data analysis as they are accustomed. If data analysis reveals the need for a union type then they must implement an interface. Here is where specification is separated from implementation. An interface contains the purpose statements and the method headers for the valid operations. If the data definition is parameterized then the interface must be parameterized. If an operation is parameterized then the method implementing it must also be parameterized.

The next step has students develop unit tests. Tests must be written for every operation for each subtype.  The tests, once again, are written before implementing the interface in order to gain insight into how to implement the operations.

After writing tests, students proceed with the implementation of the interface. They must implement the interface for each subtype in the union type in a separate class. That is, the number of classes needed is equal to the number of subtypes in the union type. Each class must implement all the operations in the interface. The implementation in each class is specific to a subtype. This step is reminiscent to students of writing functions for each subtype for functional interfaces. Finally, each class is bound to the interface using, for example, \texttt{implements} in \texttt{Java}.

Once the interface is implemented there is a good opportunity to talk to students about polymorphic dispatch and why conditionals are not required to distinguish among subtypes. After all this is accomplished, students may engage in implementing the solution to the problem. Students are presented with two choices:
\begin{enumerate}
  \item Extend the interface
  \item Write a program that uses an instance of the union type
\end{enumerate}
The first option is chosen when a new operation is needed (or requested by a client). In this case, a method solving the problem for each subtype must be added to each class. This method must also be added to the interface. The second option is chosen when the problem being solved is not a required operation for the union type. In this case a separate class is written and a conditional is required in the method to distinguish among the subtypes.

\begin{figure}[t]
\begin{alltt}
     public interface ILIST<X>
     \{ILIST<X> cons(X val); 
      // Purpose: add given X to front of this list

      boolean isEmpty();     
      // Purpose: Determine if list is empty

      X first() throws Exception; 
      // Purpose: Return the first list element

      ILIST<X> rest() throws Exception;  
      // Purpose: Return rest of this list

      public boolean equals(IList<X> l); 
      // Purpose: Is this = given list

      <Y> ILIST<Y> map(IFun<X,Y> f); 
      // Purpose: Apply given function to this

      <A> A foldl(IFun2<X, A> f, A res);
      // Purpose: Accumulate given function application from left to right
     \}
\end{alltt}
\caption{Java Interface for (listof X).} \label{jiilox}
\end{figure}

To make the steps of the design recipe concrete consider implementing \texttt{(listof X)}. Given that there is variety in the data an interface is needed. We may implement the interface as displayed in \Cref{jiilox}. It is highlighted to students how the types become part of the code. The interface is parameterized with \texttt{X} because the data definition is parameterized with \texttt{X}. The methods \texttt{map} and \texttt{foldl} are parameterized with the types, respectively \texttt{Y} and \texttt{A}, not known to \texttt{this} list. Finally, any operation that throws an error must be implemented by a method that throws an \texttt{Exception} and the types \texttt{IFUN} and \texttt{IFUN2} are wrapper interfaces to define, respectively, a one-input and two-input function.

\begin{figure}[!t]
\begin{alltt}
class ListTests2
\{@Test
  public void test()
 \{IList<Integer> N  = new MTLIST2<Integer>();
  IList<Integer> N1 = N.cons(4).cons(6).cons(3);
  IList<Integer> N2 = N.cons(4).cons(6).cons(3);
  IList<Integer> N3 = N.cons(3).cons(6).cons(4);
  IList<String>  E  = new MTLIST2<String>();
  IList<String>  L0 = E.cons("pal!").cons("there ").cons("Hi ");
  try
  \{assertEquals(N1.first(), (Integer) 3);
   assertEquals(N1.rest().equals(N.cons(4).cons(6)), true);
   assertEquals(N.isEmpty(), true);
   assertEquals(N1.isEmpty(), false);
   assertEquals(N.cons(10).equals((new mtList<Integer>()).cons(10)), true);
   assertEquals(
   N1.cons(0).equals((new mtList<Integer>()).cons(4).cons(6).cons(3).cons(0)),
   true);
   assertEquals(N.equals(N1), false);
   assertEquals(N1.equals(N2), true);
   assertEquals(E.map(s -> s.length()).equals(N), true);
   assertEquals(L0.map(s -> s.length()).equals(N1), true);
   assertEquals(N.foldl((n, r) -> r.cons(n), N).equals(N), true);
   assertEquals(N2.foldl((n, r) -> r.cons(n), N).equals(N3), true);   \}\}\}
\end{alltt}
\caption{\texttt{JUnit} Tests for \texttt{ILIST<X>}.} \label{listtests}
\end{figure}

The next step of the design recipe asks students to write unit tests for the interface. Students know that the tests for each operation must cover all the varieties in the union type. Writing tests for \texttt{ILIST<X>} provides us with the opportunity to expose students for the first time to casting and to $\lambda$-expressions in \texttt{Java}. Casting is explained in the context, for example, of letting the compiler know that we mean \texttt{Integer} 3 and not the \texttt{int} 3. The first exposure to $\lambda$-expressions is done untyped. Nonetheless, students are advised that sometimes these expressions must be typed by the programmer. One important thing to note is that \texttt{exceptions} are not tested. \Cref{listtests} displays a sample testing file developed in class following student advice. Students find the syntax verbose and awkward at times (i.e., $\lambda$-expressions) but they do not feel overwhelmed by the material given that it is introduced in a context that is familiar to them from designing using a functional language.

\begin{figure}[t]
\begin{minipage}[b]{.5\linewidth}
\begin{alltt}
class mtList<X>
       implements ILIST<X>

\{
 mtList() \{\}

 ILIST<X> cons(X v)
 \{return((new consList<X>(v,this)));\}

 X first() throws Exception
 \{throw new Exception("\dotss");\}

 ILIST<X> rest()
        throws Exception
 \{throw new Exception("\dotss");\}

 boolean isEmpty() \{return(true);\}

 boolean equals(IList<X> l)
 \{return(l.isEmpty());\}

 <Y> ILIST<Y> map(IFun<X, Y> f)
 \{return(new mtList<Y>());\}

 <A> A foldl(IFun2<X, A> f, A ac)
 \{return(ac);\} \}
\end{alltt}
\subcaption{Empty-List Class}\label{elistclass}
\end{minipage}
\begin{minipage}[b]{.5\linewidth}
\begin{alltt}
class consList<X>
       implements ILIST<X>
\{X f;  ILIST<X> r;
 consList(X v, ILIST<X> rest)
 \{f = v;  r = rest;\}

 ILIST<X> cons(X v)
 \{return((new consList<X>(v, this)));\}
 X first() \{return(f);\}
 ILIST<X> rest()\{return(r);\}
 boolean isEmpty() \{return(false);\}

 boolean equals(IList<X> l)
 \{try \{return(
    (this.first().equals(l.first())) &&
     this.rest().equals(l.rest()));\}
  catch(Exception e)
  \{System.out.println(\dotss);
   return(false); \}\}
 <R> ILIST<R> map(IFun<X, R> f)
 \{return(
   new consList<R>(f.f(this.first()),
                this.rest().map(f)));\}
 <R> R foldl(IFun2<X,R> fn, R ac)
 \{return(this.rest().foldl(
        fn, fn.f(this.first(), ac)));\}\}
\end{alltt}
\subcaption{Cons-List Class}\label{clistclass}
\end{minipage}
\caption{\texttt{ILIST<X>} Interface Implementation.}
\label{ooilox}
\end{figure}

The next step of the design recipe has students implement the interface. Students are reminded that an interface for a union type must be implemented in the form of a \texttt{class} for each subtype. Each method in \texttt{ILIST<X>} must be implemented for the \texttt{empty} and the \texttt{cons} list. This implementation is displayed in \Cref{ooilox}. The reader can appreciate how remarkably similar the functional interfaces and the classes are. There are some points that must be highlighted to students:
\begin{description}
  \item[this] The concept of \texttt{this} is introduced and substitutes references to \texttt{service-manager} in \Cref{ilox}.
  \item[Dot Composition] Messages are passed to objects in the same manner as in functional interfaces. For example, we write \texttt{((rest \textquotesingle{map}) f))} for \texttt{map} in \Cref{iclist}. We write \texttt{this.rest().map(f)} in \Cref{clistclass}.
\end{description}

The final step of the design recipe has students write code to solve a problem. This exercise is similar to \texttt{getLevel} for \texttt{Student} in \Cref{objects}. In the interest of brevity the repetition of such a development is omitted. Students end the steps of the design recipe by running the tests and, if necessary, redesigning.

\section{Abstraction}
\label{abstraction}

Program design may result in code repetition. This commonly occurs when functions are developed for similar types. One way to eliminate the repetition is to develop parameterized data definitions. Students usually face two challenges that prevent them from having repetitive code. The first is that they do not realize the significant similarities between data definitions. These result in functions that are very similar. The second is that possible similarities may not be immediately obvious to students. The first may be resolved using functional abstraction. The second may be resolved using code refactoring (a topic they have not studied before). Abstraction in an \oop \ course is used to motivate abstract classes, inheritance, and code refactoring.

\subsection{Student Background}

Students arrive to the classroom knowing how to do functional abstraction. A typical student may be presented the following functions to square a list of numbers and to extract the names of a list of inventory records (abbreviated \texttt{(listof ir)}):
\begin{alltt}
     ;; (listof number) \arrow (listof number)
     ;; Purpose: Square the given list of numbers
     (define (sqr-list L)
       (cond [(empty? L) \elist]
             [else (cons (sqr (first L)) (sqr-list (rest L)))]))

     ;; (listof ir) \arrow (listof string)
     ;; Purpose: Extract the name in the given list of ir
     (define (names a-lir)
       (cond [(empty? L) \elist]
             [else (cons (ir-name (first L)) (names (rest L)))]))
\end{alltt}
The student immediately observes that these functions are very similar and are candidates for abstraction. They perform this task by applying the design recipe for abstraction \cite{HtDP2}:
\begin{alltt}
     1. Compare functions and mark differences
     2. Define the abstract function with the differences as parameters
     3. Define and test the original functions using the abstract function
     4. Write the signature for the abstract function
\end{alltt}
Applying the design recipe results in the following code\footnote{Tests are omitted due to limited space.}:
\begin{alltt}
     ;; [X Y] (listof X) (X \arrow Y) \arrow (listof Y)
     ;; Purpose: Apply the given function to the given list's elements
     (define (mapf L f)
       (cond  [(empty? L) \elist]
              [else (cons (f (first L))  (mapf (rest L) f))]))
     ;; (listof number) \arrow (listof number)
     ;; Purpose: Square the given list of numbers
     (define (sqr-list L) (mapf L sqr))
     ;; (listof ir) \arrow (listof string)
     ;; Purpose: Extract the name in the given list of ir
     (define (names ls) (mapf L ir-name))
\end{alltt}
Students appreciate the reduction in repetition and the elegance of the resulting code (especially if the programming language provides the abstract function they discovered).

\subsection{Abstract Classes and Inheritance}

Students invariably notice repetitions and ask for abstraction. For example, they notice that in \Crefrange{elistclass}{clistclass} the \texttt{cons} method is exactly the same. This provides the opportunity to motivate abstract classes and inheritance. Students are presented with the following design recipe for abstraction:
\begin{alltt}
     1. Compare classes and mark similarities
     2. Define an abstract class containing the similarities
     3. Have classes extend abstract class to inherit similarities
     4. Refactor code to create similarities
\end{alltt}
This design recipe is strikingly similar to the design recipe for functional abstraction, which puts students at ease. In Step 1, instead of looking for differences students must look for similarities. In Step 2, the similarities are migrated to an abstract class. These similarities may include methods and instance variables of the same type. If the original classes implement an interface then the abstract class must implement the same interface. In step 3, students associate the original classes with the new abstract class. In \texttt{Java} this is done by using \texttt{extends}. That is, the original classes extend the abstract class. In this manner, all the methods and the instance variables in the abstract class are inherited and available in the original classes. It is important to explain to students that abstract classes can not be instantiated.

Applying the first three steps of the design recipe to the classes in \Crefrange{elistclass}{clistclass} yields code structured as follows:
\begin{alltt}
     abstract class AList<X> implements IList<X>
     \{IList<X> cons(X v)
      \{return(new NMTLIST<X>(v, this));\}\}

     class mtList<X> extends AList<X>
     \{ \dotss \}

     class consList<X> extends AList<X>
     \{ \dotss \}
\end{alltt}
The only changes in the classes is the migration of \texttt{cons} to the abstract class and the use of \texttt{extends} in the class headers.

\subsection{Code Refactoring}

At a first glance there are no more similarities among the classes after abstracting away \texttt{cons}. At this point students are introduced to code refactoring. Students are explained that code refactoring changes existing code without changing the semantics. In our context, refactoring is used to create similarities among classes so that they can be abstracted away. Students are explained that they may write methods in terms of other methods much like \texttt{sqr-list} and \texttt{names} above are rewritten using \texttt{mapf} (or the built-in \texttt{map}). If refactoring a method across all classes in a union type yields repetitions then they may be abstracted away.

As an example students are asked to consider the implementation of \texttt{equals} in \Crefrange{elistclass}{clistclass}. The code clearly does not look the same. We can observe, however, that if the type of \texttt{this} and the given list are different then the answer is \texttt{false}. Otherwise, \texttt{foldl} may be used to test for equality. Now that the equals method in both classes is the same it may be migrated to the abstract class to yield:
\begin{alltt}
     abstract class AList<X> implements IList<X>
     \{IList<X> cons(X v)
      \{return(new NMTLIST<X>(v, this));\}

      boolean equals(IList<X> l)
      \{if ((this.isEmpty() && !l.isEmpty()) ||
            (!this.isEmpty() && l.isEmpty()))
       \{return(false);\}
       else \{return(this.foldl((x, r) ->
                               \{try \{return(x.equals(l.first()) && r);\}
                                catch(Exception e)
                                \{System.out.println("\dotss" + e.getMessage());
                                 return(false);\}\},
             true));\}\}\}
\end{alltt}

\section{Student Feedback}
\label{feedback}

\pgfplotstableread[row sep=\\,col sep=&]{
cat    & prop \\
1 & 0 \\
2 & 0 \\
3 & 0.1875 \\
4 & 0.4375 \\
5 & 0.375 \\
}\bps

\pgfplotstableread[row sep=\\,col sep=&]{
cat    & prop \\
1 & 0 \\
2 & 0.0625 \\
3 & 0.1875 \\
4 & 0.375 \\
5 & 0.375 \\
}\emp

\pgfplotstableread[row sep=\\,col sep=&]{
cat    & prop \\
1 & 0 \\
2 & 0.0625 \\
3 & 0.125 \\
4 & 0.3125 \\
5 & 0.5 \\
}\is

\pgfplotstableread[row sep=\\,col sep=&]{
cat    & prop \\
1 & 0.133333 \\
2 & 0.133333 \\
3 & 0.4 \\
4 & 0.2 \\
5 & 0.133333 \\
}\st

\begin{figure}[t]
\begin{minipage}[b]{.5\linewidth}
\begin{tikzpicture}
    \begin{axis}[
            ybar,
            symbolic x coords={1,2,3,4,5},
            xtick=data,
        ]
        \addplot[fill=black] table[x=cat,y=prop]{\bps};
    \end{axis}
\end{tikzpicture}
\subcaption{Better Problem Solver.}\label{bpsbc}
\end{minipage}
\begin{minipage}[b]{.5\linewidth}
\begin{tikzpicture}
    \begin{axis}[
            ybar,
            symbolic x coords={1,2,3,4,5},
            xtick=data,
        ]
        \addplot[fill=black] table[x=cat,y=prop]{\emp};
    \end{axis}
\end{tikzpicture}
\subcaption{Feel Empowered.}\label{empbc}
\end{minipage}
\caption{Student Feedback I.}
\label{sf1}
\end{figure}
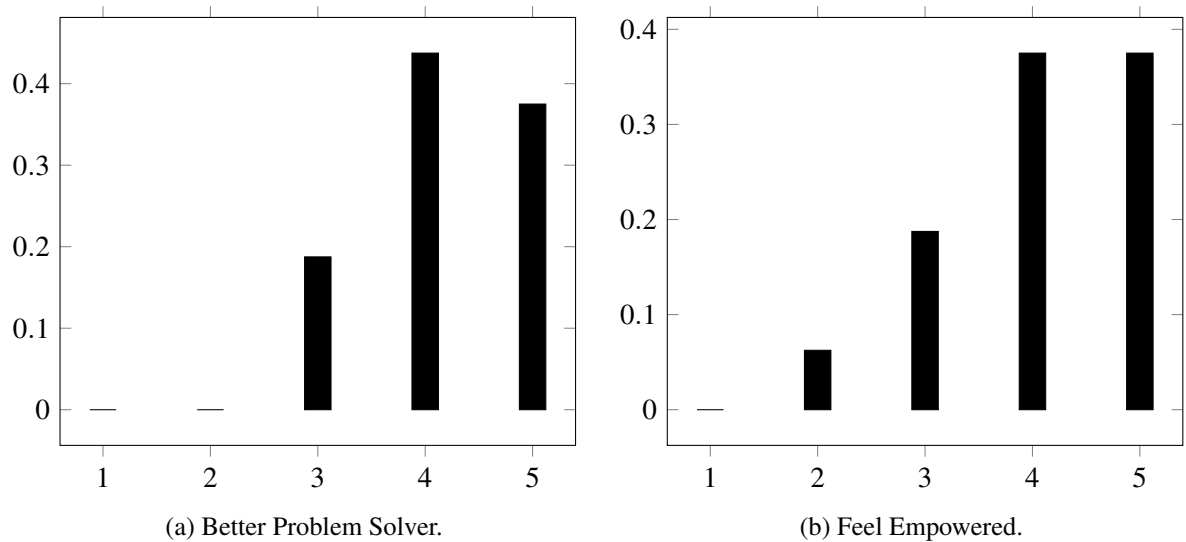

Both quantitative and qualitative data was gathered from students in the course at the end of the semester. The class consisted of 16 students (all male given that, unfortunately, we had no female enrollment in the course). The average age of the students is 19.56. All the students are Computer Science majors except one majoring in English.

Students were asked \emph{Do you feel you are a better problem solver by knowing how to design and implement \oo \ programs?} on a scale from 1 (Not at all) to 5 (Very much so). \Cref{bpsbc} summarizes the responses. We observe that all students responded positively (range 3-5) affirming that they feel they are better problem solvers. In fact, students felt quite strongly about this with 81\% of students in the 4-5 range. Closely related to this students were asked \emph{Do you feel empowered by knowing how to design and implement \oo \ programs?} on a scale from 1 (Not at all empowered) to 5 (Very much empowered). The bar chart in \Cref{empbc} summarizes the responses. We observe that students overall feel empowered with 94\% of responses in the 3-5 range. In fact, students feel quite empowered as evidenced by 75\% of responses in the 4-5 range. This data attests that students feel that they have acquired new and useful skills through the material presented.

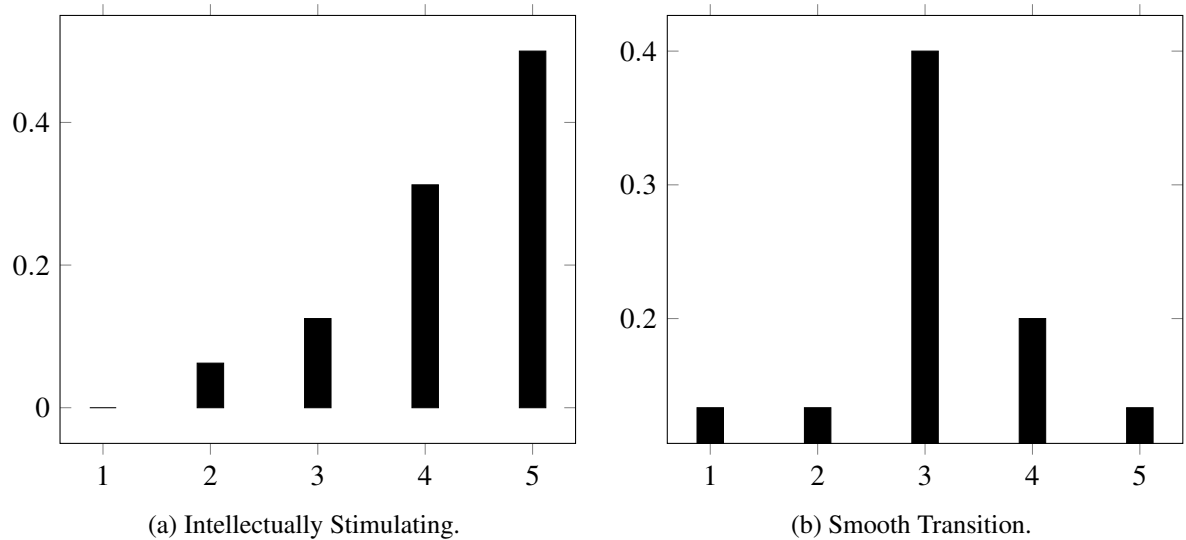
\begin{figure}[t]
\begin{minipage}[b]{.5\linewidth}
\begin{tikzpicture}
    \begin{axis}[
            ybar,
            symbolic x coords={1,2,3,4,5},
            xtick=data,
        ]
        \addplot[fill=black] table[x=cat,y=prop]{\is};
    \end{axis}
\end{tikzpicture}
\subcaption{Intellectually Stimulating.}\label{isbc}
\end{minipage}
\begin{minipage}[b]{.5\linewidth}
\begin{tikzpicture}
    \begin{axis}[
            ybar,
            symbolic x coords={1,2,3,4,5},
            xtick=data,
        ]
        \addplot[fill=black] table[x=cat,y=prop]{\st};
    \end{axis}
\end{tikzpicture}
\subcaption{Smooth Transition.}\label{stbc}
\end{minipage}
\caption{Student Feedback II.}
\label{sf2}
\end{figure}

As mentioned in \Cref{intro} a potential concern was students becoming bored by having a follow-up design-based course that relies heavily on design recipes. After all, it is possible that students may not be interested in rehashing design recipes albeit in a different programming paradigm. To ascertain students' feelings they were asked \emph{How intellectually stimulating is \oo \ programming?} on a scale from 1 (Not at all stimulating) to 5 (Very Stimulating). The responses are displayed in \Cref{isbc}. Students find the material intellectually stimulating with 94\% responding in the 3-5 range. Once again, we may observe that students in general feel very strongly as may be observed with 81\% of the responses in the 4-5 range. This strongly suggests that students are not bored by studying applications of design recipes in a new paradigm. In fact, it confirms that students need to see how the design skills they have internalized apply in a different programming context (i.e., a different programming language and paradigm).

Another concern mentioned in \Cref{intro} is whether or not this material represented a smooth transition from 2 first-year courses a la \htdp \ to an \oop \ course. To ascertain students' attitudes they were asked \emph{How smooth is the transition from the Racket student languages to Java?} on a scaled from 1 (Not at all smooth) to 5 (Extremely smooth). The responses are summarized in \Cref{stbc}. Overall, students feel the transition is smooth with 73\% of responses in the 3-5 range. Only one third of the students feel very strongly (responses in the 4-5 range) that the transition was smooth. About a quarter of the respondents (26\%) feel that the transition was not relatively smooth. The biggest problem students had was navigating \texttt{Java} error messages, which they did not find very useful. This was followed by the lack of much feedback on failed tests and the syntax of the language (especially $\lambda$-expressions).

On the qualitative side students were asked \emph{What was the worst part of the course? What did you like the least?} Sample responses are:
\begin{itemize}
  \item Learning a new syntax though it gets easier with time.
  \item Lambda functions in \texttt{Java}. They're much more difficult to understand and implement.
  \item Everything we learned I hated it all. I loved programming in AP CS, but I hate this.
\end{itemize}
These responses confirm the instructor's impressions that students did not like \texttt{Java}'s verbosity and really did not like the syntax for $\lambda$-expressions. The third response above is not typical, but does occur sometimes with students that took a \texttt{Java} programming course in high school. Some of these students expect to be given code to edit or programming problems that are tightly-coupled with an example done in class. Asking them to design the solution to a problem from scratch does not fulfill their expectations and they require more attention to help them overcome their frustration. Helpful techniques to address their needs include encouraging them to work in groups and to attend tutoring. Rather surprisingly, it is worth noting that students did not complain or raise concerns about having a \texttt{Java} textbook that is tightly-coupled with design-based programming. Students do inquire about such a textbook when they get stuck, but this shortcoming was not elevated to the level of a complaint in this group of students.

Students were also asked \emph{What was the best part of this course? What did you like best?}. Sample responses are:
\begin{itemize}
  \item Abstraction was awesome.  Interfaces were cool.  Creating objects was cool.
  \item Learning to implement our knowledge into object oriented design.
  \item Solving problems with code.
\end{itemize}
The first response above is common in spirit among other responses given. It points to students truly feeling they understand the technology (i.e., programming constructs) that they are using. The second response is also, in spirit, similar to other responses. Students welcomed being taught how their design experience is applicable to program development using a popular language in industry. The third comment above is not uncommon. In general, students find problem solving fun when they have steps (i.e., a design recipe) to guide them.

\section{Concluding Remarks and Future Work}
\label{fw}
This article presents part of the methodology developed to introduce students to some abstractions and programming constructs available in an object-oriented programming language. In particular, it discusses how to introduce students to classes, interfaces, generic programming, inheritance, polymorphic dispatch, $\lambda$-expressions, and abstract classes. Although the approach was developed to provide a smooth transition from a design-based introduction to programming using functional programming to design-based object-oriented programming, the methodology has proven successful with transfer students (i.e., students that transfer from other universities without the benefit of a designed-based course in their background). The data collected from students to date indicates that the approach is well-received and that students find the mentioned transition to be smooth.

Future work includes developing a methodology to introduce students to modifiers that restrict access to methods and instance variables, to Iterators, and to subtype specializations. In addition, future work includes developing a methodology to introduce students to design patterns as a mechanism to abstract away frequent steps \oo \ programmers perform during software development.

\bibliographystyle{eptcs}
\bibliography{TFPIE-2021}

\end{document}